\documentclass[aps,twocolumn,showpacs]{revtex4-1}
\usepackage{latexsym}
\usepackage{graphics}
\usepackage{graphicx,amssymb,epsfig,epsf,amsmath}

\def\be{\begin{equation}}
\def\lan{\left\langle}
\def\ran{\right\rangle}
\def\ee{\end{equation}}
\def\barr{\begin{array}}
	\def\earr{\end{array}}

\def\nn8{\\}

\def\ed{\end{document}}

\def\cm{{\bf m}}
\def\cs{{\bf s}}

\begin{document}
\title{Distribution of Higher Order Spacing Ratios in Interacting Many Particle Systems}
\author{Priyanka Rao$^1$, Manan Vyas$^2$, N. D. Chavda$^1$}
\affiliation{$^1$Department of Applied Physics, Faculty of Technology and Engineering, The Maharaja Sayajirao University of Baroda, Vadodara-390001, India \\$^2$Instituto de Ciencias F{\'i}sicas, Universidad Nacional Aut{\'o}noma de M\'{e}xico, 62210 Cuernavaca, M\'{e}xico % etc
}                     % Do not remove
\begin{abstract}
We study the distribution of non-overlapping spacing ratios of higher-orders for complex interacting many-body quantum systems, with and without spin degree of freedom (in addition to the particle number). The Hamiltonian of such systems is well represented by embedded one- plus two-body random matrix ensembles (with and without spin degree of freedom) for fermionic as well as bosonic systems. We obtain a very good correspondence between the numerical results and a recently proposed generalized Wigner surmise like scaling relation. These results confirm that the proposed scaling relation is universal in understanding spacing ratios in complex many-body quantum systems. Using spin ensembles, we demonstrate that the higher order spacing ratio distributions can also reveal quantitative information about the underlying symmetry structure.
\end{abstract}

\maketitle
\section{Introduction}
\label{sec:1}

The study of spectral fluctuations is very crucial for understanding the inherent complexities of complex quantum systems. The  spectral fluctuation measures, used in many different fields, are modeled through Random Matrix Theory (RMT) \cite{porter65,Mehta2004,Akemann2011}. These are useful in characterizing distinct phases observed in physical systems such as localized or delocalized phase \cite{geraedts2016},  insulating or metallic phase of many-body systems \cite{hasegawa2000,nishigaki99}, integrable or chaotic limit of the underlying classical system \cite{reichl2013}, and low-lying shell model or mixing regime of nuclear spectra \cite{weiden2009,brody81}. It is now well established that a quantum system is chaotic if its spectral properties follow one of the three classical ensembles, the Gaussian orthogonal (GOE), unitary (GUE) or symplectic (GSE) ensemble depending on the symmetries of the Hamiltonian \cite{haake2010}.

The nearest neighbor spacing distribution (NNSD), $P(s)ds$, giving degree of level repulsion is one of the most popular measures in the study of spectral fluctuations. For time reversal and rotational invariant systems (represented by GOE), as conjectured by Bohigas et al \cite{Bohigas} and proved for certain systems by Haake et al \cite{Ha-07}, if a quantum system is chaotic, NNSD follows the Wigner surmise, $P(s) = (\pi/2) s \exp (-\pi s^2/4)$, which indicates the presence of `level repulsion'. However, as established by Berry and Tabor~\cite{Berry}, if a quantum system is integrable, NNSD follows Poisson distribution [$P(s) = \exp (-s)$] displaying `level clustering'.

For a given set of energy levels (or eigenvalues), construction of NNSD requires `unfolding' of the spectra in order to remove the secular variation in the density of eigenvalues~\cite{brody81,haake2010}. This is a cumbersome and non-unique numerical procedure. Also, for many-body systems such as Bose-Hubbard model, unfolding procedure of the spectra becomes non-trivial as the density of states is not a smooth function of energy in the strong interaction domain~\cite{Huse2007,koll2010,Coll2012}. Moreover, there are discrepancies between spectral and ensemble unfoldings for non-ergodic random matrices.

In the past, Oganesyan and Huse \cite{Huse2007} introduced the distribution $P(r)$ of the ratio of consecutive level spacings of the energy levels which does not require unfolding as it is independent of the form of the density of the energy levels. Importantly, Atas et. al \cite{ABGR-2013} derived expressions for $P(r)$ for the classical GOE, GUE and GSE ensembles of random matrices. The statistics of ratio of spacings has been used to quantify the distance from integrability on finite size lattices \cite{koll2010,Coll2012}, to investigate many-body localization \cite{Huse2007,OPH2009,Pal-10,Iyer-12}, to establish that finite many particle quantum systems, modeled by embedded random matrix ensembles, with strong enough interactions follow GOE \cite{CK}, to study spectral correlations in diffused van der Waals clusters \cite{HBCK} and to analyze spectra of uncorrelated random graph network \cite{Ru-19}. Recently, exact distribution of spacing ratios for random and localized states in quantum chaotic systems is obtained using a $3\times3$ random matrix model \cite{tekur2018-1} and a generalized form of Wigner surmise has been proposed for the distribution of non-overlapping spacing ratios of higher-orders  \cite{tekur2018-2}. An extension of the Wigner surmise for distribution of higher order spacing ratios was also proposed in the past~\cite{porter63,abul99} and applied to systems with mixed regular-chaotic dynamics~\cite{abul2000}.

The spectral fluctuations, obtained using the discrete levels drawn from the same subspace, of complex quantum systems are known to follow RMT. For mixed spectra, the levels from different symmetries are superposed, resulting in level clustering, which gives misleading results~\cite{Berry}. Using the higher order spacing statistics, one can obtain information about symmetry structure for any arbitrary sequence of measured or computed levels without desymmetrization~\cite{tekur2018-3}. This method is straightforward compared to the complicated and approximate methods based on two-level cluster function for a composite spectrum~\cite{Guhr1990,levi-86}.

In a large number of investigations carried out during last two decades, it is well established that embedded Gaussian orthogonal ensembles \cite{French-Wong,Bohigas-Flores} of one plus two-body interactions [EGOE(1+2)] apply in a generic way to isolated finite interacting many-particle quantum systems such as nuclei, atoms, quantum dots, small metallic grains, interacting spin systems modeling quantum computing core and so on \cite{vkbk01,Weiden,Gomez,Manan-thesis,VKBK2014}. Recently these models have also been used successfully in understanding high energy physics related problems. Random matrix models with two-body interactions [EGOE(2)] among complex fermions are known as complex Sachdev-Ye-Kitaev models in this area \cite{Davison-PRB-2017,Bul-JHEP-2017,Rosenhaus-JPA-2019}. In the present work, we analyze generic properties of non-overlapping higher order spacing ratios for several embedded ensembles, both for fermionic and bosonic systems, with and without spin degrees of freedom. We also show that the quantitative information about the symmetry structure of the system can be obtained using higher order spacing ratios for embedded ensembles with spin degree of freedom.

The paper is organized as follows. Analytical results for the probability distribution of the ratio of consecutive level spacings and higher order spacing ratios for GOE are briefly discussed in Section~\ref{sec:2}. The five different EGOEs, used in the present analysis, are defined in Section~\ref{sec:3}. Numerical results of the distribution of higher order ratio of consecutive spacings and related averages are presented in Section~\ref{sec:4}. Finally, Section~\ref{sec:5} gives the concluding remarks.

\section{Probability distribution of higher order spacing ratios}
\label{sec:2}

Understanding and deriving generic results for fluctuation properties has widespread applications in all branches of physics, mathematics, engineering and so on \cite{Akemann2011,RMT-books}. Consider an ordered set of eigenvalues (energy levels) $e_n$, where $n = 1,2,...d$. The consecutive eigenvalue spacings are given by $s_n = e_{n+1} - e_n$ and the ratios of two nearest neighbor or consecutive eigenvalue spacings are $r_n=s_{n+1}/s_n$. Using an exact calculation for $3 \times 3$ Gaussian random matrices, the probability distribution $P(r)$ for consecutive eigenvalue spacings for GOE is derived to be given by \cite{ABGR-2013},
\begin{equation}\label{eq:1}
P(r) = \frac{27}{8} \frac{(r+r^2)}{(1+r+r^2)^{5/2}} \;.
\end{equation}
Nearest neighbor spacing ratios $r$ probe fluctuations in spectral scales of the order of unit mean spacing. Many different variants of consecutive level spacing ratios have been studied recently \cite{CK,tekur2018-1,CDK2014}

The non-overlapping higher order spacing ratios can be defined as \cite{tekur2018-2},
\begin{equation}\label{eq:2}
		r^{(k)}_n = \frac{s^{(k)}_{n+k}}{s^{(k)}_n} = \frac{e_{n+2k} - e_{n+k}}{e_{n+k} - e_{n}}; \;\; n,k = 1,2,3...
\end{equation}
such that there is no shared eigenvalue spacing in the numerator and denominator. Higher order spacing ratios $r^{(k)}$ probe fluctuations in spectral interval of $k$ mean spacings. Denoting the non-overlapping $k$-th order probability distribution by $P^k(r)$, there exists a scaling relation between $P^k(r)$ and $P_\alpha(r)$ for the class of Wigner-Dyson random matrices \cite{tekur2018-3},
\be
\begin{array}{rcl}\label{eq:3}
P^k(r) &=& P_\alpha (r) \\
 &=& C_\alpha \; \displaystyle\frac{(r+r^2)^\alpha}{(1+r+r^2)^{1 + 3\alpha/2}} \;;\\
\alpha &=& \displaystyle\frac{(k+2)(k+1)}{2} - 2,\;\; k \ge 1 \;.
\end{array}
\ee
Here, $C_\alpha$ is the normalization constant. The modified parameter $\alpha \geq 4$ can take large integer values and it accounts for the dependence on order $k$ of the spacing ratio. This scaling relation holds good for Gaussian and circular ensembles of random matrix theory and for several physical systems such as spin chains, chaotic billiards, Floquet systems and measured nuclear resonances \cite{tekur2018-2}.

We analyze $P^k(r)$ for embedded ensembles for fermion and boson systems with and without spin degree of freedom and show that the functional form of $P^k(r)$ is generically identical to $P_\alpha(r)$ for complex many-body quantum systems.

\section{Embedded ensembles for fermion and boson systems with and without spin degree of freedom}
\label{sec:3}

In this section, we define the models that we use to represent complex many-body interacting quantum systems. Embedded Gaussian Orthogonal Ensembles (EGOE) are random matrix models with two-body interactions among its constituents (fermions or bosons) that model Hamiltonians $H$ of interacting many-body quantum systems \cite{brody81,French-Wong,Bohigas-Flores,Manan-thesis,VKBK2014}. Given a number $m$ ($m > 2$) of particles (fermions or bosons), they are distributed among $N$ number of single particle (sp) states. As the particles are in an average field generated by other particles, it is appropriate to add a mean-field term $h(1)$ to the Hamiltonian. Thus, with random two-body interactions $V(2)$, the model EGOE(1+2) is defined by the Hamiltonian,
\be
H = h(1) + \lambda \{V (2)\} \;,
\label{eq.egoe1}
\ee
with $\lambda$ being the strength of the interactions. Here, notation $\{\;\}$ denotes an ensemble. The $V(2)$ matrix is chosen to be a GOE in two-particle spaces and the one-body Hamiltonian $h(1) = \sum_i \epsilon_i n_i$ is specified by sp energies $\epsilon_i = i + 1/i$. Here, $n_i$ are number operators acting on sp states $i = 1, 2, \ldots, N$. Distributing these $m$ particles in $N$ sp levels generates the $d$-dimensional many-particle basis. Action of Hamiltonian $H$ on these many-particle basis states generates EGOE(1+2). Without loss of generality, we choose the average spacing between the sp levels to be unity so that $\lambda$ is unit-less.

When we have fermions as constituents, $d= {N \choose m}$ and the two-body matrix elements, chosen to be from GOE, are properly anti-symmetrized. These models are called EGOE(1+2) for spinless fermions. When we have bosons as constituents, $d = {N-m+1 \choose m}$ and two-body matrix elements, chosen to be from GOE, are symmetrized. These models are called BEGOE(1+2).

In order to analyze universal properties of systems with spin degree of freedom, it is important to include spin $\cs$ as an additional degree of freedom in these models. Given $m$ fermions distributed in $\Omega$ number of sp orbitals each with spin $\cs = 1/2$,  the number of sp states is $N = 2 \Omega$. As two-particle spin $\cs$ can take two values (0 and 1),  $V(2) = V^{s=0}(2) \oplus V^{s=1}(2)$, that is, $V(2)$ is a direct-sum matrix of matrices in spin 0 and 1 spaces, chosen to be independent GOEs, with respective dimensions $\Omega(\Omega+1)/2$ and $\Omega(\Omega-1)/2$. The many-particle spin $S$ can take values $m/2, m/2-1, \ldots, 0$ or $1/2$. Thus, EGOE(1+2)-$\cs$ \cite{KCS-2006,mkc-2010} is defined by the Hamiltonian $H$ in Eq. \eqref{eq.egoe1} with $V(2) = V^{s=0}(2) \oplus V^{s=1}(2)$. The many-particle Hamiltonian matrices are first constructed in smallest spin projection basis ($M_S^{min}$) using spinless formulation and then the states with a given $S$ value are projected using the $S^2$ operator. Many-particle Hamiltonian matrices have a block diagonal structure with each block corresponding to an embedded ensemble with a given total spin $S$. Similarly, for two species boson systems, it is possible to consider bosons with a fictitious spin ($F=1/2$) degree of freedom. Then, we have BEGOE(1+2)-$F$ \cite{vyas-12} defined by Hamiltonian $H$ in Eq. \eqref{eq.egoe1} with $V(2) = V^{f=0}(2) \oplus V^{f=1}(2)$; $F = m/2, m/2-1, \ldots, 0$ or $1/2$.

Usually, one associates integer spins with bosonic systems and therefore, we have also analyzed boson systems with spin-one degree of freedom, denoted by BEGOE(1+2)-$S1$ \cite{Deota}. Consider $m$ bosons distributed in $\Omega$ orbitals each with spin $1$. Here, $N = 3 \Omega$ and the random interaction $V(2)$ will
be of the form $V(2) = V^{s=0}(2) \oplus V^{s=1}(2) \oplus V^{s=2}(2)$ as the two-particle spins are $s = 0$,
$1$ and $2$. Here, $V^{s=0}(2)$, $V^{s=1}(2)$ and $V^{s=2}(2)$ are chosen to be independent GOEs in two-particle spaces. Thus, BEGOE(1+2)-S1 is defined by Hamiltonian $H$ in Eq. \eqref{eq.egoe1} with $V(2) = V^{s=0}(2) \oplus V^{s=1}(2) \oplus V^{s=2}(2)$.

In the present paper, we make the following choices to analyze spacing ratios:

\begin{enumerate}
	
\item EGOE(1+2) with $m = 6$ and $N = 12$ resulting in $H$ matrix dimension $d = 924$. We choose $\lambda = 0.1$.
	
\item EGOE(1+2)-{\cs} with $m = 6$, $\Omega = 8$, $S = 0-3$ with $H$ matrix dimensions $1176$, $1512$, $420$ and $28$ respectively. We choose $\lambda = 0.1$.
	
\item BEGOE(1+2) with $m = 10$ and $N = 5$ resulting in $H$ matrix dimension $d=1001$. We choose $\lambda = 0.03$.
	
\item BEGOE(1+2)-$F$ with $m = 10$, $\Omega = 4$, and $F = 0-5$ with $H$ matrix dimensions $196$, $540$, $750$, $770$, $594$ and $286$ respectively. We choose $\lambda = 0.05$.
	
\item BEGOE(1+2)-$S1$ with $m = 8$, $\Omega = 4$, $S = 0-8$ with $H$ matrix dimensions $714$, $1260$, $2100$, $1855$, $1841$, $1144$, $840$, $315$ and $165$ respectively. We choose $\lambda = 0.2$.

\end{enumerate}

In all the cases, we construct 500 member ensemble and have chosen interaction strength to be sufficiently large so that there is sufficient mixing among the basis states and the system is in the Gaussian domain, i.e. the eigenvalue density and local density of states (LDOS) will be close to Gaussian and level and strength fluctuations (using an appropriate unfolding function) will follow GOE  \cite{Manan-thesis,VKBK2014,Deota}. Now, we will present numerical results.

\section{Numerical results}
\label{sec:4}

\begin{figure}
\begin{center}
\includegraphics[width=\linewidth]{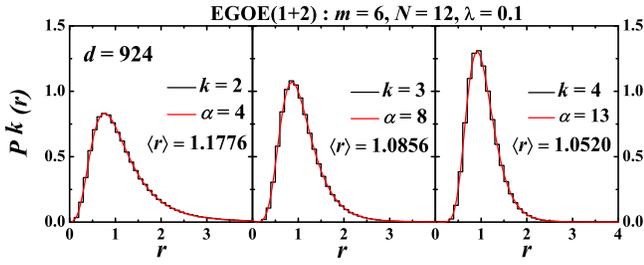}
\end{center}
\caption{Histograms represent probability distribution of the $k$-th order spacing ratios $r$ (represented by $P^k(r)$) for a 500 member EGOE(1+2)
ensemble with $k = 2$, $3$, and $4$. The red smoothed curves (represented by $P_{\alpha}(r)$) are obtained using Eq.~\eqref{eq:3} with $\alpha$ values as mentioned in each panel.}
\label{fig:1}
\end{figure}

\begin{figure}
\centering
\includegraphics[width=\linewidth] {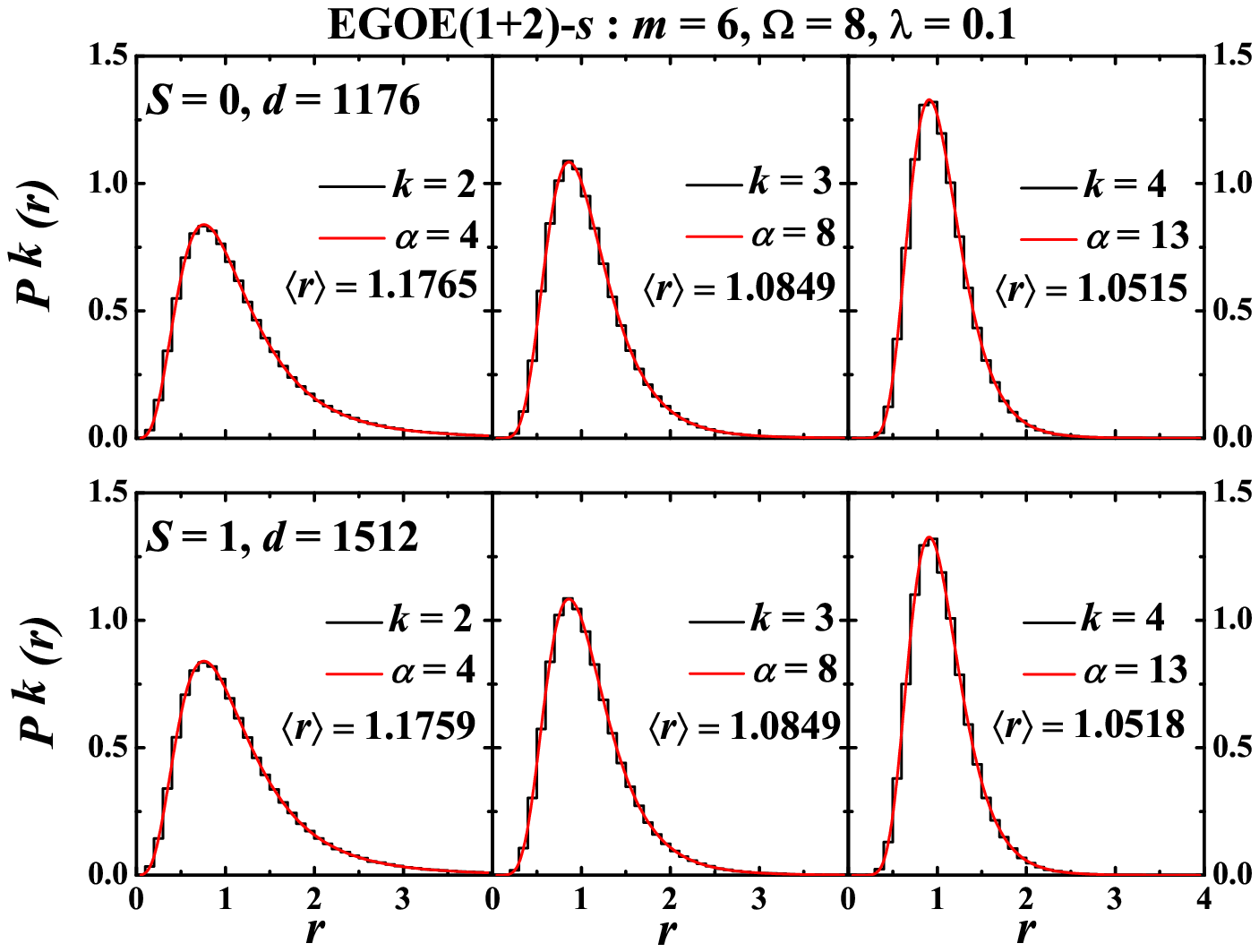}
\caption{Same as figure \ref{fig:1} but results are for a 500 member EGOE(1+2)-$\cs$ ensemble. Top panel corresponds to $S = 0$ and bottom panel corresponds to $S = 1$.}
	\label{fig:2}       % Give a unique label
\end{figure}

\begin{figure}
\centering
\includegraphics[width=\linewidth]{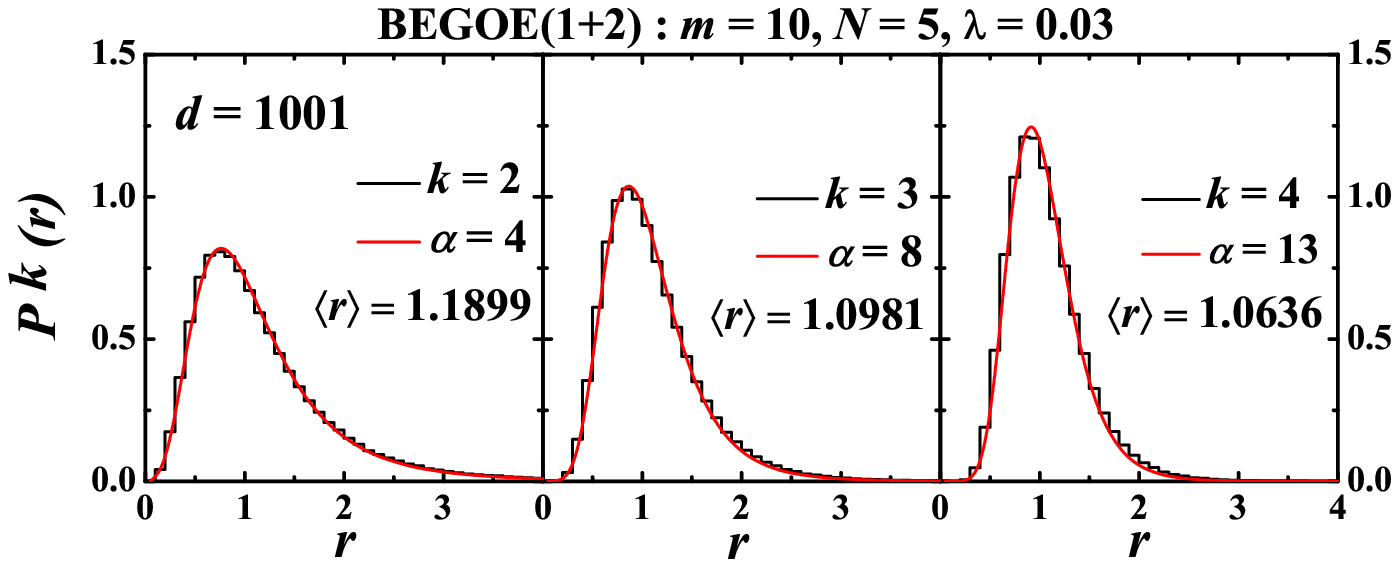}
\caption{Same as figure \ref{fig:1} but results are for a 500 member BEGOE(1+2) ensemble.}
	\label{fig:3}       % Give a unique label
\end{figure}

\begin{figure}
\centering
			\includegraphics[width=\linewidth] {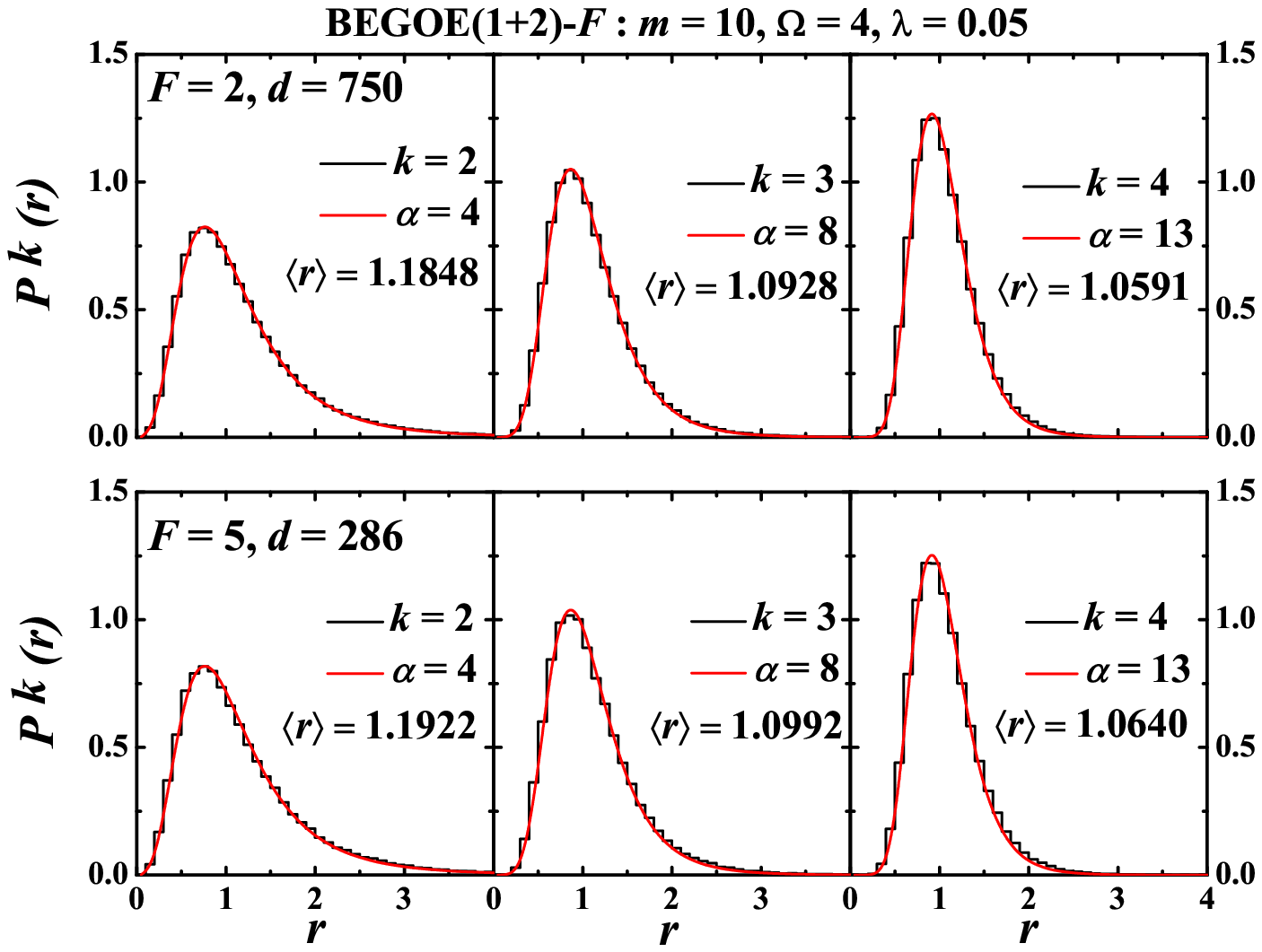}
	\caption{Same as figure \ref{fig:1} but results are for a 500 member BEGOE(1+2)-$F$ ensemble. Top panel corresponds to $F = 2$ and bottom panel corresponds to $F = 5$.}
	\label{fig:4}       % Give a unique label
\end{figure}

\begin{figure}
\centering
			\includegraphics[width=\linewidth] {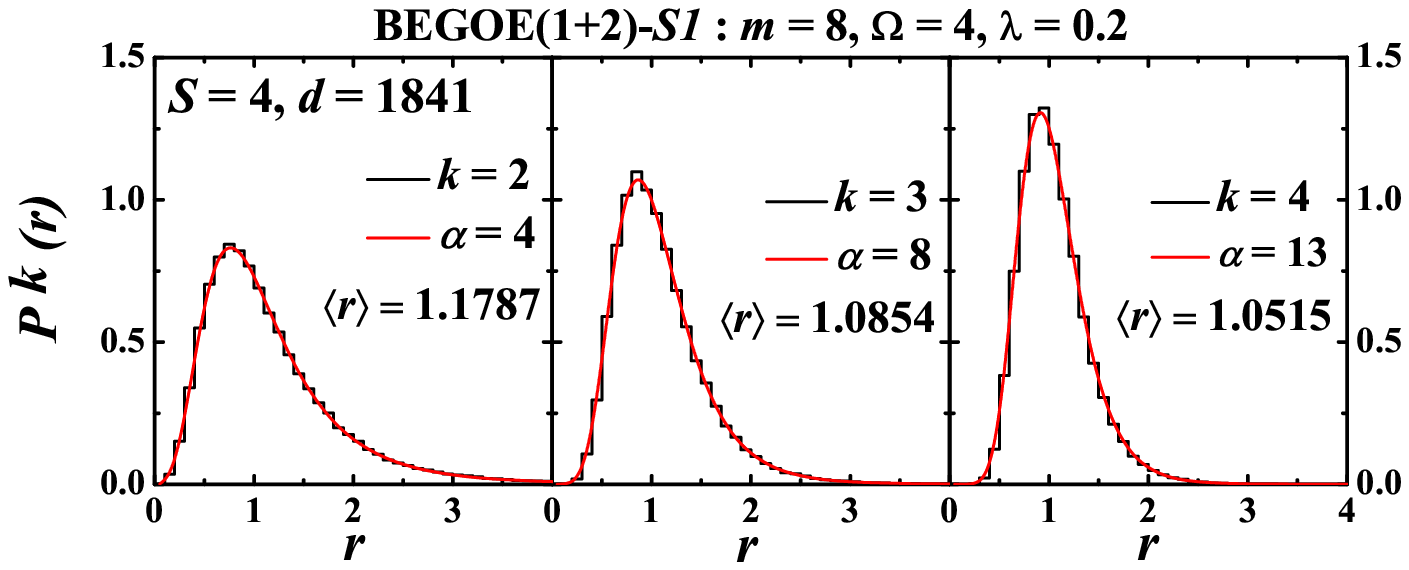}
		\caption{Same as figure \ref{fig:1} but results are for a 500 member BEGOE(1+2)-$S1$ ensemble with spin $S = 4$.}
	\label{fig:5}       % Give a unique label
\end{figure}

Using the definition given in Section \ref{sec:2}, we have constructed  $k$-th order spacing ratio distribution $P^k(r)$ for the EGOE(1+2) models defined in Section \ref{sec:3}. Using central 80\% of the spectrum, we construct numerical histograms for $P^k(r)$ with a bin-size of 0.1 and $k = 2$, 3 and 4. Figure \ref{fig:1} shows the spacing ratio distribution $P^k(r)$ (black histogram) for EGOE(1+2) compared with $P_\alpha(r)$ (smooth red curves). Here the $\alpha$ values are 4, 8 and 13 for $k = $  2, 3 and 4 respectively. The average values of spacing ratios ${\lan r \ran}$ are as given in the figure.  Using Eq. \eqref{eq:3}, the average values of spacing ratios ${\lan r \ran}_{th}$ are 1.1747, 1.0855 and 1.0521 respectively for $k = 2$, $3$ and $4$. Similarly, Figures \ref{fig:2} - \ref{fig:5} show variation in $P^k(r)$ compared with $P_\alpha(r)$ respectively for EGOE(1+2)-$\cs$, BEGOE(1+2), BEGOE(1+2)-$F$ and BEGOE(1+2)-$S1$. For all the examples considered, we find that ${\lan r \ran}_{EE} \sim {\lan r \ran}_{th}$. We also obtain good agreement when we include all the levels in the analysis, unlike for nearest neighbor spacing distribution which also gets affected by the choice of unfolding function. As seen from these figures, we obtain excellent agreement between numerical histograms and $P_\alpha(r)$ establishing that Eq. \eqref{eq:3} explains the universal features in variation of higher order spacing ratios in many-body interacting quantum systems, with and without spin degree of freedom.

\begin{figure}
\centering
			\includegraphics[width=\linewidth] {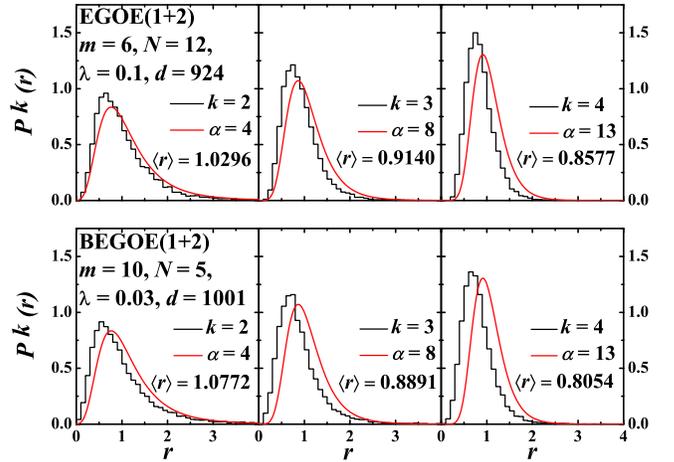}
	\caption{Probability distribution $P^k(r)$ (histograms) of the $k$-th order spacing ratios $r$ for the lowest 20 energy levels using EGOE(1+2) (top panel) and BEGOE(1+2) (bottom panel) ensembles. The red smoothed curves are obtained using Eq.~\eqref{eq:3} with $\alpha$ values as mentioned in each panel.}
	\label{fig:6}       % Give a unique label
\end{figure}

Energy levels of EGOE(1+2) close to the ground state (tails of the energy spectrum) generate large fluctuations compared to that of GOE fluctuations \cite{flores2001}. In order to test the validity of Eq. \eqref{eq:3} close to the ground state, we analyzed spacing ratio distributions $P^k(r)$ using the lowest $20$ energy levels for EGOE(1+2) and BEGOE(1+2) ensembles (with the choice of parameters as outlined in Sec. \ref{sec:3}). The numerical histograms for these are compared with $P_\alpha(r)$ (smooth red curves) in Fig. \ref{fig:6}. Numerical results show a clear deviation from the trend predicted by Eq. \eqref{eq:3}, with deviation increasing with increasing $k$. Also, ${\lan r \ran}$ values are found to be smaller than the corresponding ${\lan r \ran}_{th}$ values. Therefore, although one need not exclude the spectrum tails while analyzing non-overlapping spacing ratios, Eq. \eqref{eq:3} does not explain the variation in spacing ratios close to the ground state.

\begin{figure}
\centering
			\includegraphics[width=\linewidth] {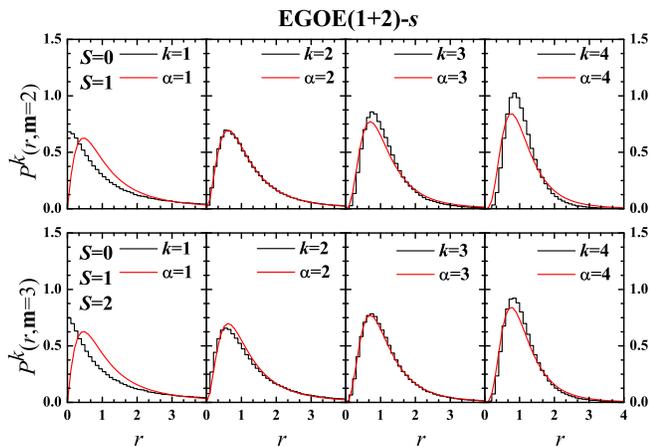}
	\caption{Histograms represent probability distribution of non--overlapping $k$-th order spacing ratios $r$ of $\cm$ independent superposed spin blocks (represented by $P^k(r,\cm)$) for a 500 member EGOE(1+2)-$\cs$ ensemble. Upper panel shows results for $\cm=2$, with spins $S=0-1$, while lower panel shows results for $\cm=3$, with spins $S=0-2$. The histograms are compared with the red smoothed curves obtained using Eq.~\eqref{eq:3} with $\alpha$ values as mentioned in each panel.}
	\label{fig:7}       % Give a unique label	
\end{figure}

\begin{figure}[tbh]
\centering
			\includegraphics[width=\linewidth] {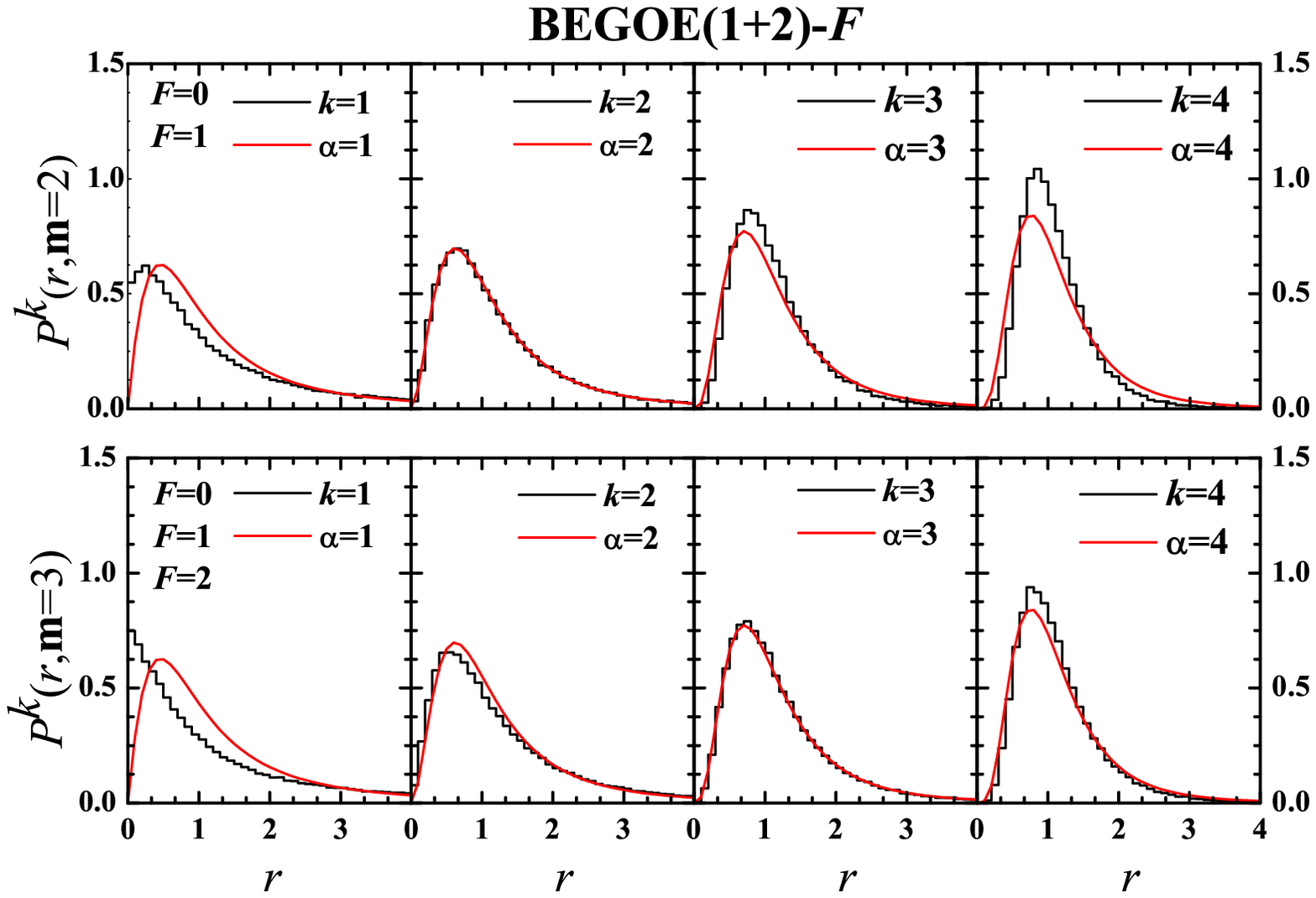}
	\caption{Same as figure \ref{fig:7} but results are for $\cm$ superposed spectra of a 500 member BEGOE(1+2)-$F$ ensemble. Upper panel shows results for $\cm=2$ with spins $F=0-1$ while lower panel shows results for $\cm=3$ with spins $F=0-2$. The histograms are compared with the red smoothed curves obtained using Eq.~\eqref{eq:3} with $\alpha$ values as mentioned in each panel.}
	\label{fig:8}       % Give a unique label
\end{figure}

\begin{figure}[tbh]
\centering
		\includegraphics[width=\linewidth] {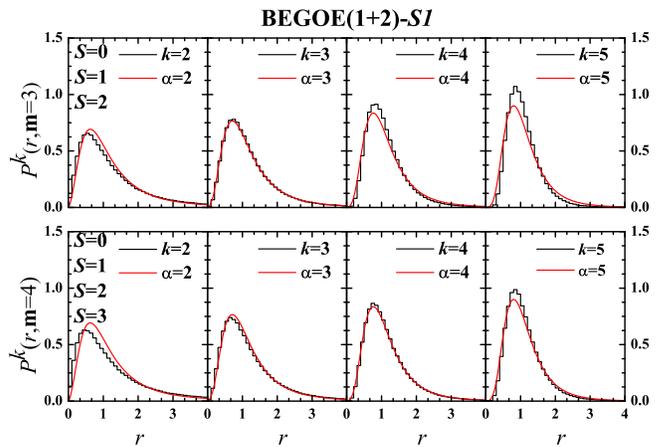}
	\caption{Same as figure \ref{fig:7} but results are for $\cm$ superposed spectra of a 500 member BEGOE(1+2)-$S1$ ensemble. Upper panel shows results for $\cm=3$ with spins $S=0-2$ while lower panel shows results for $\cm=4$ with spins $S=0-3$. The histograms are compared with the red smoothed curves obtained using Eq.~\eqref{eq:3} with $\alpha$ values as mentioned in each panel.}
	\label{fig:9}       % Give a unique label
\end{figure}

Distributions of higher-order spacing ratios can also be used to understand quantitative information regarding underlying symmetry structure in addition to explaining universal features of fluctuation characteristics. As conjectured by Dyson \cite{dyson1962} and proved by Gunson \cite{gunson1962}, the spectral statistics of two superposed circular orthogonal ensemble (COE) spectra converge to that of circular unitary ensemble (CUE).This is expected to be echoed in the distribution of level spacings and spacing ratios as well. Using examples of superposed GOE spectra, billiards, spin-$1/2$ chains and neutron resonance data, it has been demonstrated that distribution of higher order spacing ratios carry symmetry information \cite{tekur2018-3}.

As discussed in Sec \ref{sec:3}, the EGOE(1+2) models with spin, EGOE(1+2)-$\cs$, BEGOE(1+2)-$F$ and BEGOE(1+2)-$S1$ have specific spin structure: for EGOE(1+2)-$\cs$ and BEGOE(1+2)-$F$, the random interaction matrix $V(2)$ in two-particle spaces is a direct sum of matrices in spin 0 and 1 channels; and for BEGOE(1+2)-$S1$, the $V(2)$ matrix in two-particle spaces is a direct sum of matrices in spin 0, 1 and 2 channels. The many-particle Hamiltonian matrix is a block diagonal matrix with each block corresponding to EGOE(1+2) with a given spin $S$. In order to investigate whether $P^k(r)$ carries signatures of these spin structures, we superpose $\cm$ independent spin blocks and compare non-overlapping $k$-th order spacing ratios distribution $P^k(r,\cm)$ (here $P^k(r)=P^k(r,\cm=1)$) with $P_\alpha(r)$ given by Eq.~\eqref{eq:3}.

\begin{figure}[tbh]
\centering
%\begin{tabular}{c}
			\includegraphics[width=0.8\linewidth] {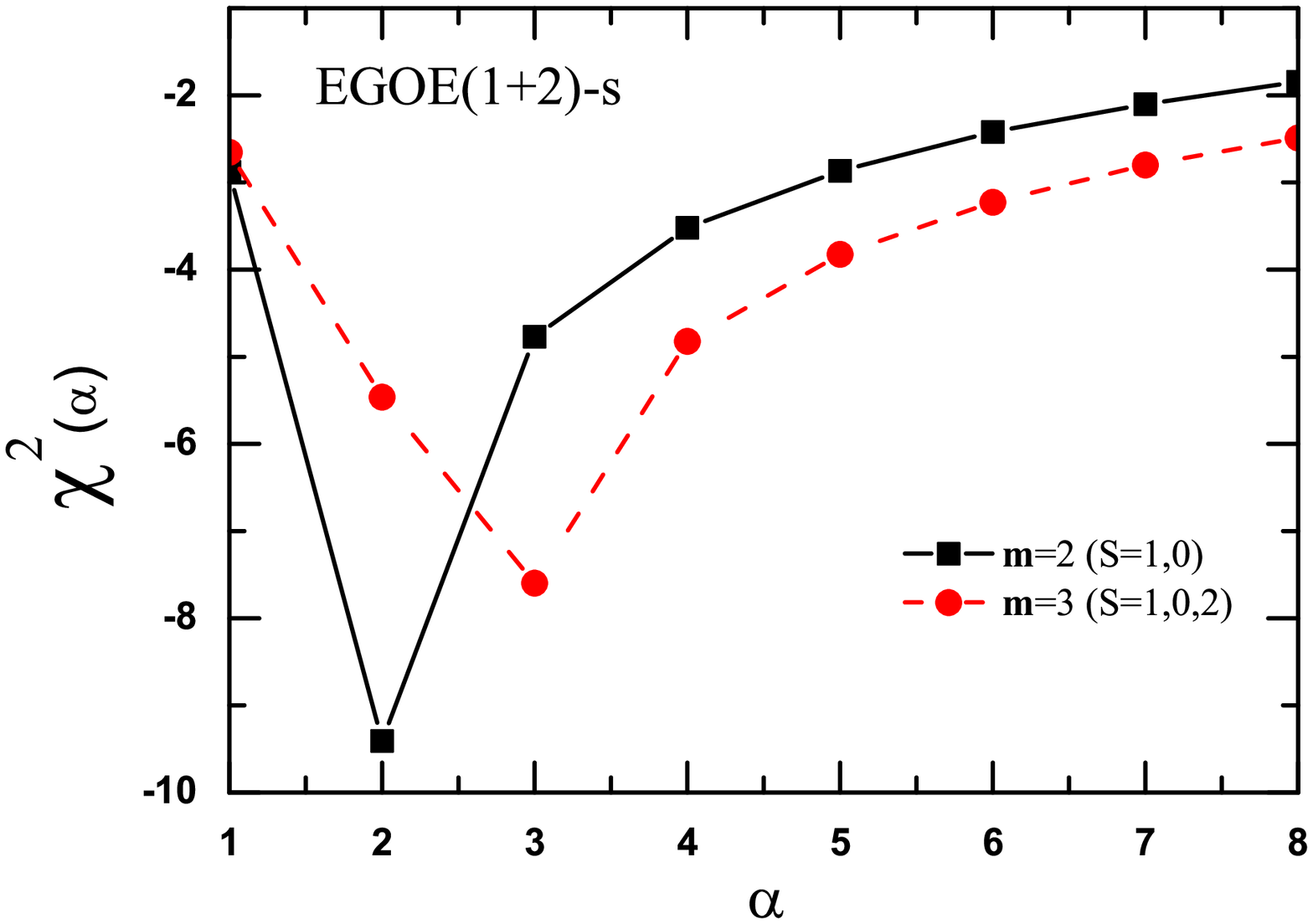}\\
			\includegraphics[width=0.8\linewidth] {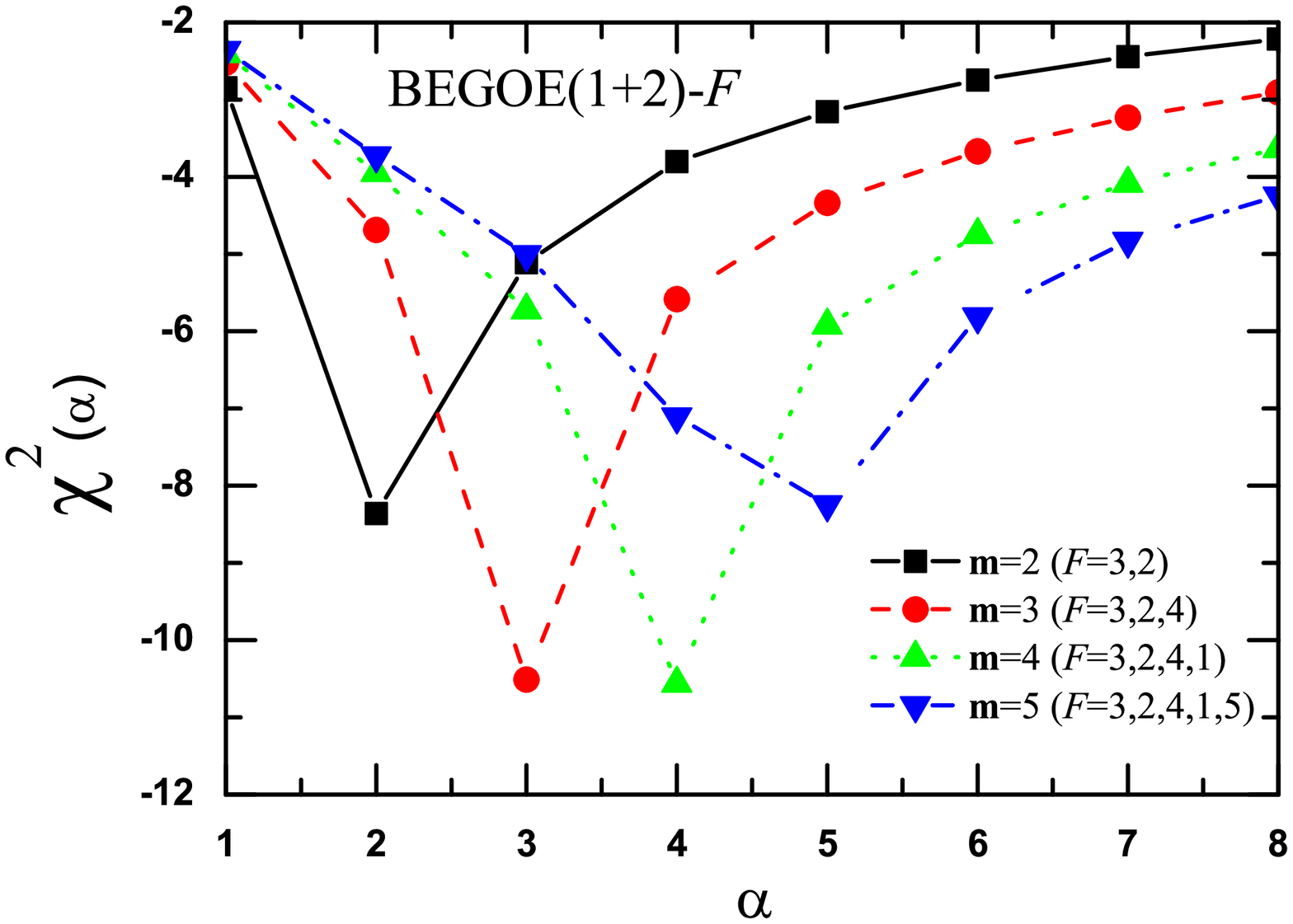}
% \end{tabular}
	\caption{Variation in $\chi^2(\alpha)$ vs. $\alpha$ for EGOE(1+2)-$\cs$ ensemble (upper panel) and  BEGOE(1+2)-$F$ ensemble (lower panel) for different $\cm$ values as mentioned in the panel. The minimum values of $\chi^2(\alpha)$ indicates  $P^k(r,\cm) \sim P_\alpha(r)$ for $\cm$ superimposed spectra.}
	\label{fig:10}       % Give a unique label	
\end{figure}

\begin{figure*}
\centering
%\begin{tabular}{c}
			\includegraphics[width=0.3\linewidth] {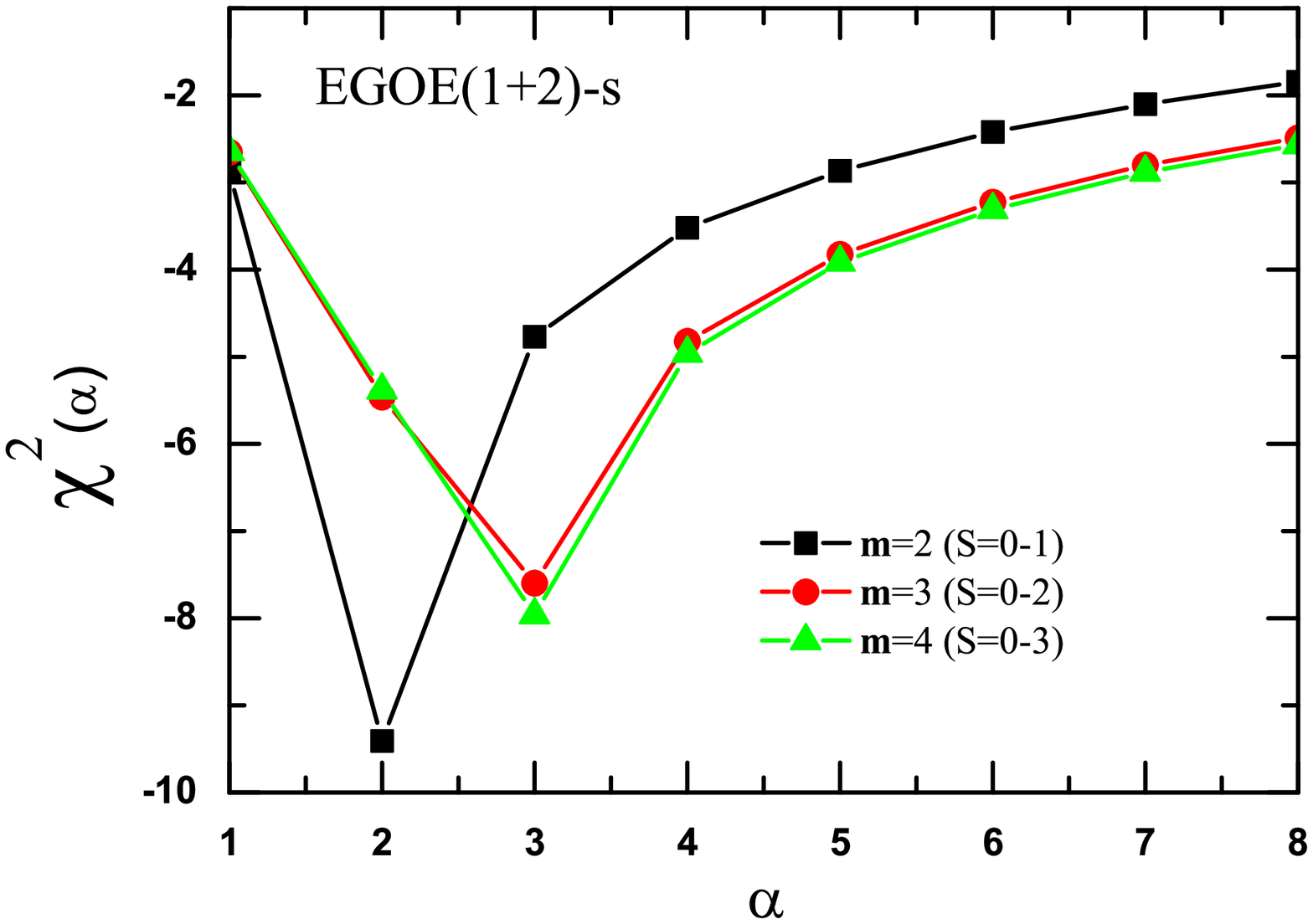}
			\includegraphics[width=0.3\linewidth] {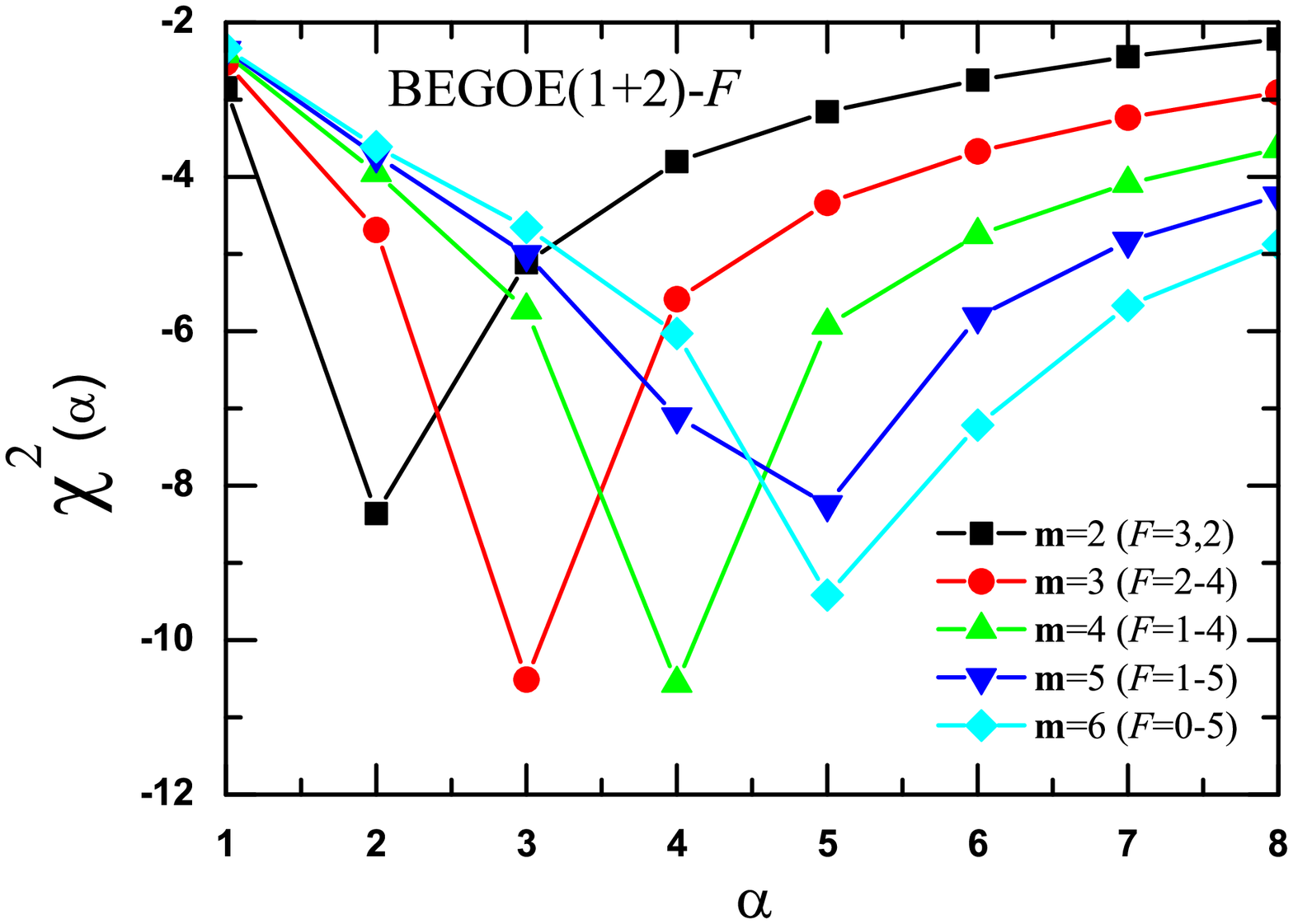}
                           \includegraphics[width=0.3\linewidth] {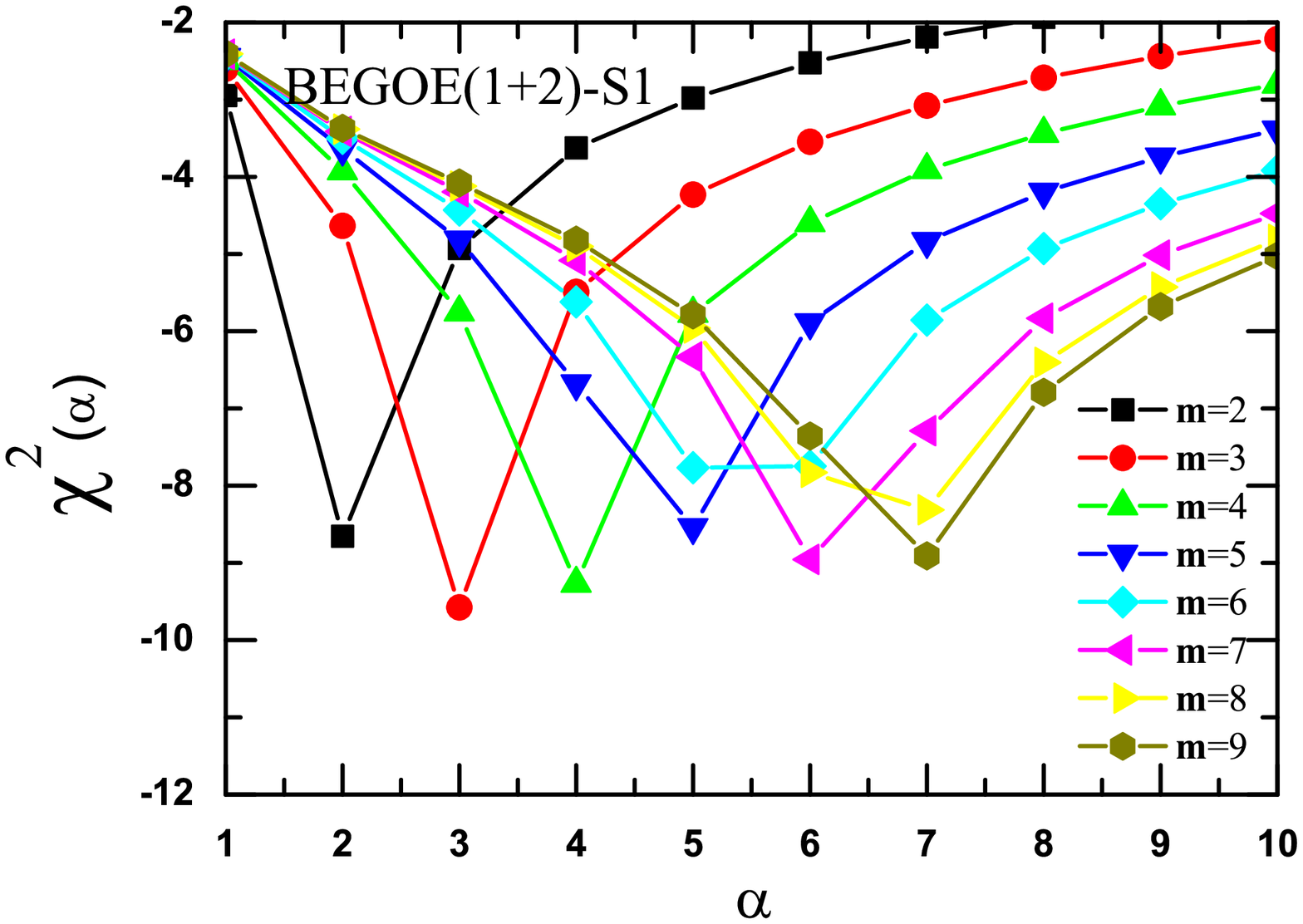}\\
			\includegraphics[width=0.3\linewidth] {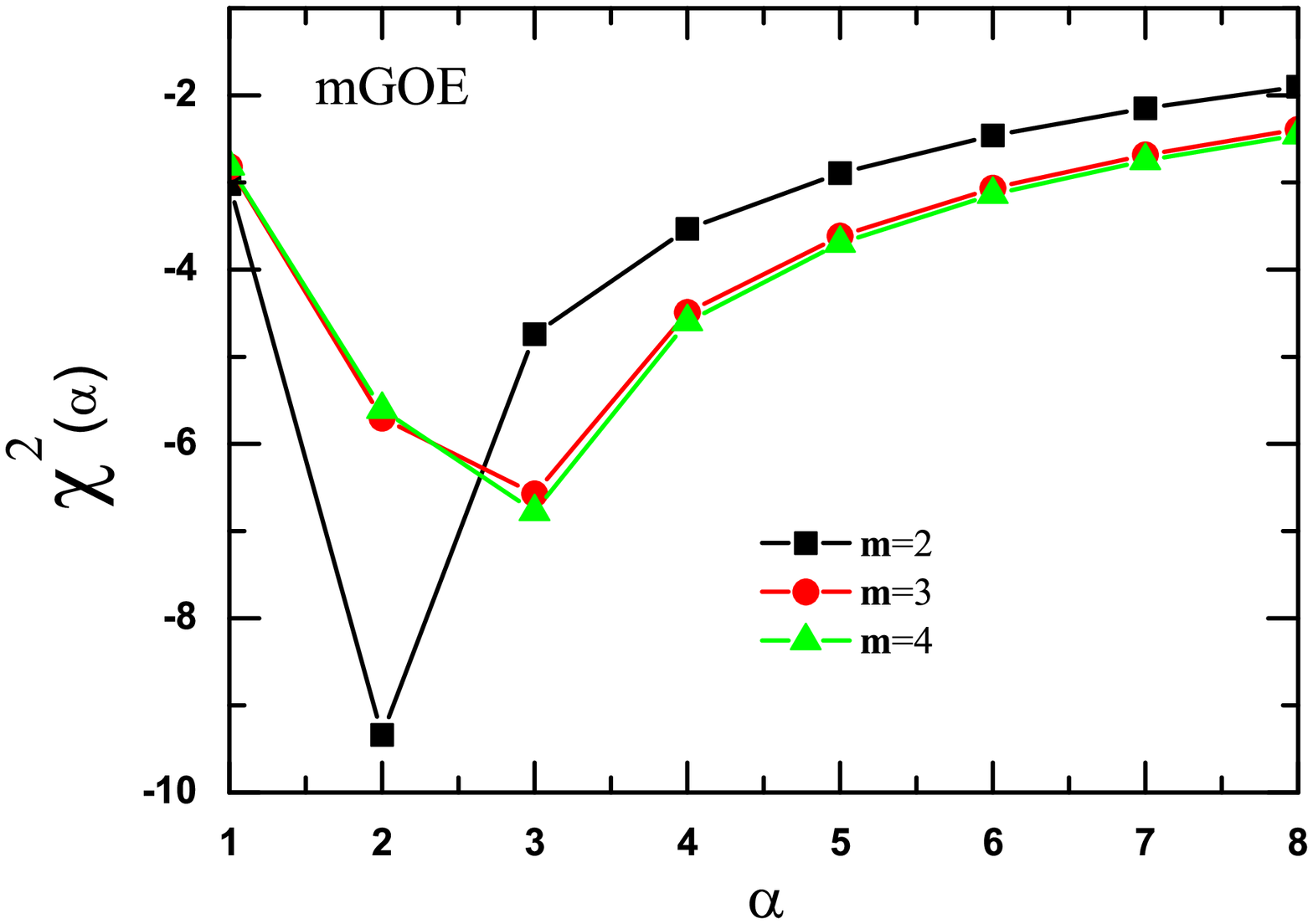}
			\includegraphics[width=0.3\linewidth] {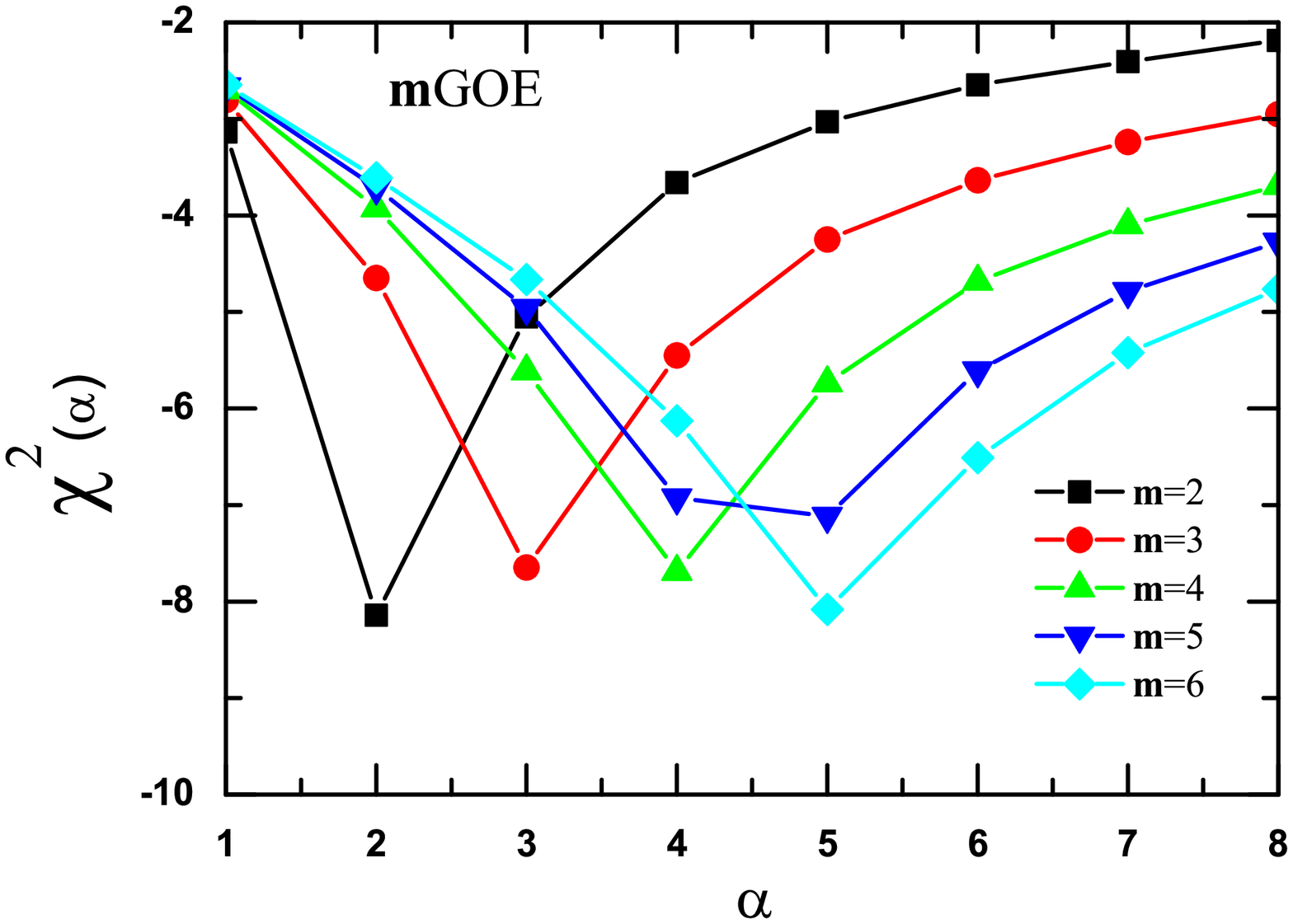}
                           \includegraphics[width=0.3\linewidth] {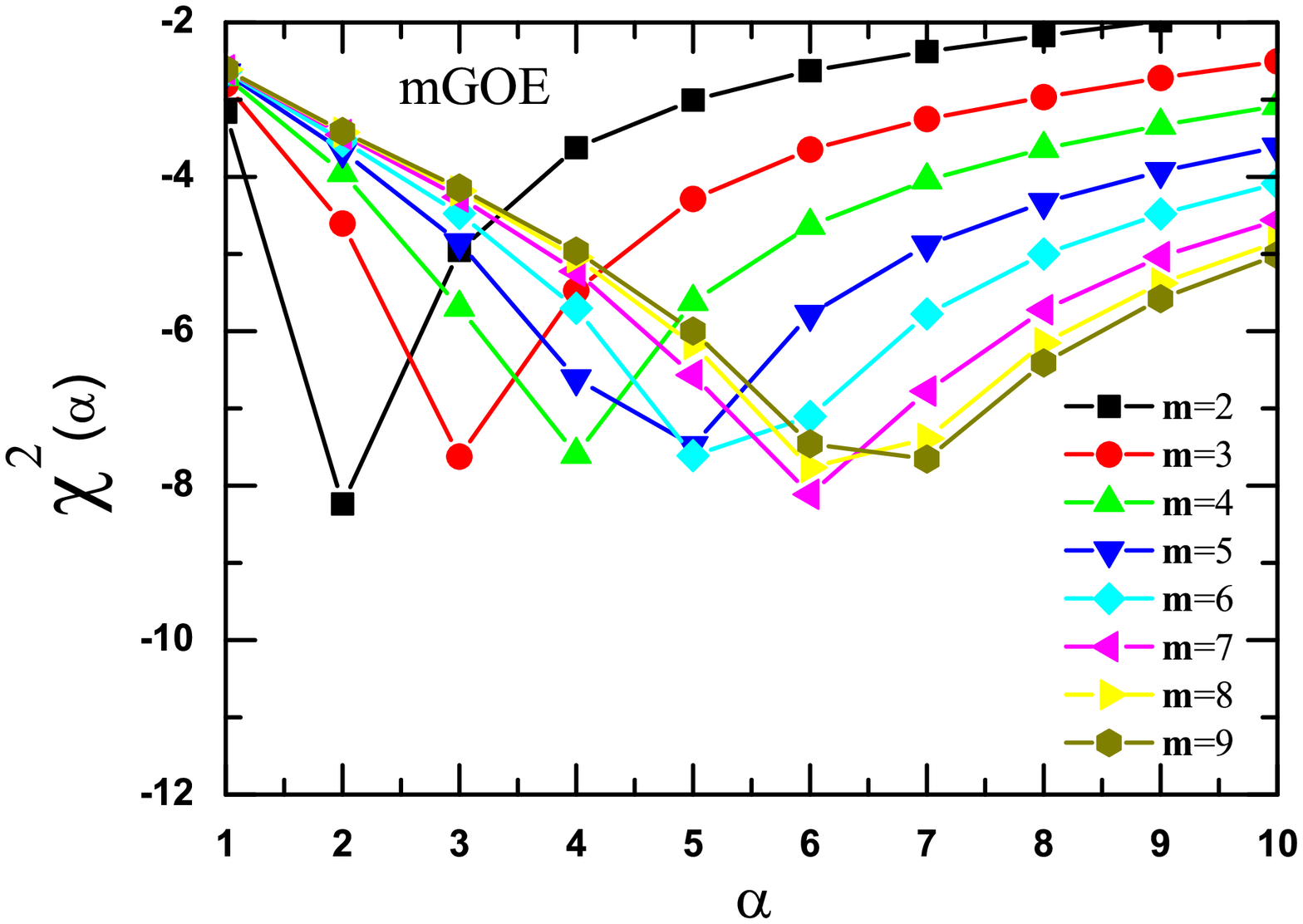}\\
% \end{tabular}
	\caption{Top panel shows variation in $\chi^2(\alpha)$ vs. $\alpha$ for EGOE(1+2)-$\cs$ (left panel), BEGOE(1+2)-$F$ (middle panel) and BEGOE(1+2)-$S1$ (right panel) for different $\cm$ values as indicated in the panel. Lower panel represents variation in $\chi^2(\alpha)$ vs. $\alpha$ for superposed $\cm$ GOE spectra of exactly same dimensions corresponding to results in the top panel for EGOE(1+2)-$\cs$, BEGOE(1+2)-$F$ and BEGOE(1+2)-$S1$ respectively. The minimum value of $\chi^2(\alpha)$ indicates  $P^k(r,\cm) \sim P_\alpha(r)$ for $\cm$ superimposed spectra.}
	\label{fig:11}       % Give a unique label	
\end{figure*}

For the choice of parameters outlined in Sec. \ref{sec:3}, results are shown in Figures \ref{fig:7}-\ref{fig:9} respectively for EGOE(1+2)-$\cs$, BEGOE(1+2)-$F$ and BEGOE(1+2)-$S1$. In these figures, the ensemble averaged histograms for $P^k(r,\cm)$ are obtained by arranging the spectra of $\cm$ spin blocks in ascending order for each member of the ensemble. Then, for each $k$, ensemble average is computed and plotted as a histogram with bin-size of $0.1$. Figure \ref{fig:7} shows $P^k(r,\cm)$ with $\cm = 2$ (upper panel) obtained by superposing spectra corresponding to $S = 0$ and $S = 1$ and $\cm = 3$ (lower panel) obtained by superposing spectra corresponding to $S = 0$, $S = 1$ and $S = 2$. The smooth (red) curves are for $P_\alpha(r)$ obtained using Eq. \eqref{eq:3}. As seen from these figures, we obtain very good agreement between numerics and theory for $\alpha = k = \cm$. There are clear deviations for all other values. These confirm the presence of $\cm$ symmetries. Similarly, Figures \ref{fig:8} and \ref{fig:9} also show excellent agreements between theory and numerics for $\alpha = k = \cm$ confirming the presence of $\cm$ symmetries.

In order to obtain the best quantitative estimate for $\alpha$, we calculate $\chi^2$ measure defined as,
\begin{equation}\label{eq:4}
\chi^2(\alpha)= \log \left\{ \int_0^\infty dr (P^{k}(r,\cm)-P_{\alpha}(r))^2 \right\} \;.
\end{equation}
Here, minimum value of $\chi^2$ indicates $P^k(r,\cm) \sim P_\alpha(r)$.
Figure \ref{fig:10} shows the variation in $\chi^2(\alpha)$ as a function of $\alpha$ for various $\cm$ values. The upper panel gives the results for EGOE(1+2)-$\cs$ and the lower panel gives the results for BEGOE(1+2)-$F$. We have not included spectra of maximum spin $S = S_{max}$ for EGOE (1+2)-$\cs$ and spectra of minimum spin $F = F_{min}$ for BEGOE(1+2)-$F$ due to small matrix dimensions. The minimum value for $\chi^2$ is obtained at $\alpha=k=\cm$, which is in agreement with results in Figures \ref{fig:7} and \ref{fig:8}. We have also confirmed this result with other combinations of superposed spectra corresponding to different spin sectors. Thus, higher order spacing ratios are useful in extracting symmetry information.

There are deviations from obtaining minimum for $\chi^2(\alpha)$ at $\alpha = k = \cm$ when the dimension of a given spin block is very small. Figure \ref{fig:11} shows variation in $\chi^2(\alpha)$ as a function of $\alpha$ for EGOE(1+2)-$\cs$ (top left panel), BEGOE(1+2)-$F$ (top middle panel) and BEGOE(1+2)-$S1$ (top right panel). Results are shown for various $\cm$ values. For EGOE (1+2)-$\cs$, the minimum for $\chi^2(\alpha)$ is not at
$\alpha = k = \cm$ for $\cm = 4$ as it is obtained by superposing four spin blocks corresponding to $S = 0-3$. Here, $S = 3$ is the maximum allowed spin and has the smallest dimension ($28$ compared to dimensions $1176$, $1512$, $420$ respectively for spins $0$, $1$ and $2$). Similarly, deviations are seen in minimum for $\chi^2(\alpha)$ from $\alpha = k = \cm$ at $\cm = 6$ for BEGOE(1+2)-$F$ and for $\cm = 6-9$ for BEGOE(1+2)-$S1$. Going further, we superposed $\cm$ GOE spectra of exact same dimensions corresponding to EGOE(1+2)-$\cs$ (top left panel), BEGOE(1+2)-$F$ (top middle panel) and BEGOE(1+2)-$S1$ (top right panel) and the results are shown in the bottom panel of the figure. These also show similar deviations in minimum for $\chi^2(\alpha)$ confirming that there are finite-size effects.

\section{Conclusion}
\label{sec:5}

We have analyzed higher order spacing ratios for interacting many-body quantum systems with and without spin degree of freedom. These complex systems are modeled by EGOE(1+2) for fermionic and bosonic systems with and without spin degree of freedom. We obtain excellent agreement between numerical results for higher order spacing ratios and Wigner surmise like scaling relation. Thus, this scaling relation is universal to understand higher order spacing ratios in complex many-body quantum systems (fermionic as well as bosonic) with rotational and time-reversal invariance, with and without spin degree of freedom. We have shown that the higher order spacing ratio distributions can also reveal quantitative information about underlying symmetry structure. Hence, the analysis of higher order spacing ratios is not only useful in studying spectral fluctuations but also reveals quantitative information about symmetry structure of complex quantum systems. Although not shown explicitly, these results are expected to extend to the unitary versions EGUE(1+2) (for both fermionic and bosonic systems, with and without spin degrees of freedom) as well. It would be interesting to analyze complex spacing ratios \cite{Prosen-2019} to characterize integrable and chaotic dynamics for the embedded unitary ensembles with various symmetries like spin, parity, total angular momentum etc. \cite{Manan-thesis,VKBK2014} and characterize universality of transition to chaos \cite{Rel-2019}.

\section*{Acknowledgement}
Authors thank V. K. B. Kota for many useful discussions. P. R. and N. D. C. acknowledge support from Department of Science and Technology(DST), Government of India [Project No.: EMR/2016/001327]. M. V. acknowledges financial support from UNAM/DGAPA/PAPIIT research grant~IA101719.


\begin{thebibliography}{99}

\bibitem{porter65} C. E. Porter, \textit{Statistical Theories of Spectra: Fluctuations} (Academic Press, New York, 1965).

\bibitem{Mehta2004} M. L. Mehta, \textit{Random Matrices} (Elsevier B.V., Netherlands, 2004).

\bibitem{Akemann2011} G. Akemann, J. Baik, P. Di Francesco (Eds.), \textit{The Oxford Handbook of Random Matrix Theory} (Oxford University Press, Oxford, 2011).

\bibitem{geraedts2016} S. D. Geraedts, R. Nandkishore, N. Regnault, Phys. Rev. B {\bf 93}, 174202 (2016).

\bibitem{hasegawa2000} H. Hasegawa, Y. Sakamoto, Progress of Theoretical Physics Supplement {\bf 139}, 112 (2000).

\bibitem{nishigaki99} S. M. Nishigaki, Phys. Rev. E {\bf 59}, 2853 (1999).

\bibitem{reichl2013} L. Reichl, \textit{The Transition to Chaos: Conservative Classical Systems and Quantum Manifestations} (Springer Science and Business Media, 2013).

\bibitem{weiden2009} H. A. Weidenm\"uller, G. E. Mitchell, Rev. Mod. Phys. {\bf 81}, 539 (2009).

\bibitem{brody81} T. A. Brody, J. Flores, J. B. French, P. A. Mello, A. Pandey, S. S. Wong, Rev. Mod. Phys. {\bf 53}, 385 (1981).

\bibitem{haake2010} F. Haake, \textit{Quantum Signatures of Chaos} (Springer-Verlag, Heidel-berg, 2010).

\bibitem{Bohigas} O. Bohigas, M.-J. Giannoni, C. Schmit, Phys. Rev. Lett. {\bf 52}, 1 (1984).

\bibitem{Ha-07} S. Heusler, S. M\"{u}ller, A. Altland, P. Braun, F. Haake, Phys. Rev. Lett. {\bf 98}, 044103 (2007).

\bibitem{Berry}  M. V. Berry, M. Tabor, Proc. Roy. Soc. (London)  {\bf A356},  375 (1977).

\bibitem{Huse2007} V. Oganesyan, D. A. Huse, Phys. Rev. B  {\bf 75}, 155111 (2007).

\bibitem{koll2010} C. Kollath, G. Roux, G. Biroli, A. M. L\"{a}uchli, J. Stat. Mech. P08011 (2010).

\bibitem{Coll2012} M. Collura, H. Aufderheide, G. Roux, D. Karevski, Phys. Rev. A {\bf 86}, 013615 (2012).

\bibitem{ABGR-2013} Y. Y. Atas, E. Bogomolny, O. Giraud, G. Roux, Phys. Rev. Lett. {\bf 110}, 084101 (2013).

\bibitem{OPH2009} V. Oganesyan, A. Pal, D. A. Huse, Phys. Rev. B {\bf 80}, 115104 (2009).

\bibitem{Pal-10} A. Pal, D. A. Huse, Phys. Rev. B {\bf 82}, 174411 (2010).

\bibitem{Iyer-12} S. Iyer, V. Oganesyan, G. Refael, D. A. Huse, Phys. Rev. B {\bf 87}, 134202 (2013).

\bibitem{CK} N. D. Chavda, V. K. B. Kota, Phys. Lett. A {\bf 377}, 3009 (2013).

\bibitem{HBCK} S.K. Haldar, B. Chakrabarti, N.D. Chavda, T.K. Das, S. Canuto, V. K. B. Kota, Phys. Rev. A {\bf 89},  043607 (2014).

\bibitem{Ru-19} G. Torres-Vargas, R. Fossion, J. A. M\'endez-Berm\'udez, Physica A {\bf 2019}, 123298 (2019).

\bibitem{tekur2018-1} S. Harshini Tekur, Santosh Kumar, M. S. Santhanam, Phys. Rev. E {\bf 97}, 062212 (2018).

\bibitem{tekur2018-2} S. Harshini Tekur, Udaysinh T. Bhosale, M. S. Santhanam, Phys. Rev. B {\bf 98}, 104305 (2018).

\bibitem{porter63} P. B. Kahn, C. E. Porter, Nucl. Phys. {\bf 48}, 385 (1963).

\bibitem{abul99} A. Y. Abul-Magd, M. H. Simbel, Phys. Rev. E {\bf 60}, 5371 (1999).

\bibitem{abul2000} A. Y. Abul-Magd, M. H. Simbel, Phys. Rev. E {\bf 62}, 4792 (2000).

\bibitem{tekur2018-3} S. Harshini Tekur, M. S. Santhanam, arXiv:1808.08541.

\bibitem{Guhr1990} T. Guhr, H.A. Weidenm\"{u}ller, Chem. Phys. {\bf 146}, 21 (1990).

\bibitem{levi-86} L. Leviandier, M. Lombardi, R. Jost, J. P. Pique, Phys. Rev. Lett. {\bf 56}, 2449 (1986).

\bibitem{French-Wong} J. B. French, S. S. M. Wong, Phys. Lett. {\bf B 33}, 449  (1970).

\bibitem{Bohigas-Flores} O. Bohigas, J. Flores, Phys. Lett. {\bf B 34}, 261 (1971).

\bibitem{vkbk01} V. K. B. Kota, Phys. Rep. {\bf 347}, 223 (2001).

\bibitem{Weiden} H. A. Weidenm\"{u}ller, G. E. Mitchell, Rev. Mod. Phys. {\bf 81}, 539 (2009).

\bibitem{Gomez} J. M. G. Gomez, K. Kar, V. K. B. Kota, R. A. Molina, A. Rela\~no, J. Retamosa, Phys. Rep. {\bf 499}, 103 (2011).

\bibitem{Manan-thesis} Manan Vyas, Ph.D. Thesis (M.S. University of Baroda, Vadodara, India, 2011), arXiv:1710.0833 (2017).

\bibitem{VKBK2014} V. K. B. Kota, \textit{Embedded Random Matrix Ensembles in Quantum Physics} (Springer-Verlag, Heidelberg, 2014).

\bibitem{Davison-PRB-2017} R. A. Davison {\it et. al.}, Phys. Rev. B {\bf 95}, 155131 (2017).

\bibitem{Bul-JHEP-2017} K. J. Bulycheva, J. High Energ. Phys. {\bf 12}, 069 (2017).

\bibitem{Rosenhaus-JPA-2019} V. Rosenhaus, J. Phys. A {\bf 52}, 323001 (2019).

\bibitem{RMT-books} R. N. Mantegna, H. E. Stanley, {\it An Introduction to Econophysics: Correlations and Complexity in Finance}, (Cambridge University Press, New York, 1999); M. Wright, R. Weiver, {\it New Directions in Linear Acoustics and Vibrations: Quantum Chaos, Random Matrix Theory and Complexity}, (Cambridge University Press, New York, 2010); Z. Bai, J. W. Silverstein, {\it Spectral Analysis of Large Dimensional Random Matrices}, Second edition (Springer, New York, 2010); P. J. Forrester, {\it Log-Gases and Random Matrices}, (Princeton University Press, USA, 2010); R. Couillet, M. Debbah, {\it Random Matrix Methods for Wireless Communications}, (Cambridge University Press, New York, 2012).

\bibitem{CDK2014} N. D. Chavda, H. N. Deota, V. K. B. Kota, Phys. Lett. A {\bf 378}, 3012 (2014).

\bibitem{KCS-2006} V. K. B. Kota, N. D. Chavda, R. Sahu, Phys. Lett. A {\bf 359}, 381 (2006).

\bibitem{mkc-2010} Manan Vyas, V. K. B. Kota, N. D. Chavda, Phys. Rev. E {\bf 81}, 036212 (2010).

\bibitem{vyas-12} Manan Vyas, N. D. Chavda, V.K.B. Kota, V. Potbhare, J. Phys. A: Math. Theor. {\bf 45}, 265203 (2012).

\bibitem{Deota} H. N. Deota, N. D. Chavda, V. K. B. Kota, V. Potbhare, Manan Vyas, Phys. Rev. E {\bf 88}, 022130 (2013).

\bibitem{flores2001} J. Flores, M. Horoi, M. M\"{u}ller, T.H. Seligman, Phys. Rev. E {\bf 63}, 026204 (2001).

\bibitem{dyson1962} F. J. Dyson, J. Math. Phys. {\bf 3}, 140 (1962); {\bf 3}, 157 (1962); {\bf 3}, 166 (1962).

\bibitem{gunson1962} J. Gunson, J. Math. Phys. {\bf 3}, 752 (1962).

\bibitem{Prosen-2019} L. S\'a, P. Ribeiro, T. Prosen, arXiv:1910.12784.

\bibitem{Rel-2019} A. L. Corps, A. Rela\~no, arXiv:1910.01434.

\end{thebibliography}
\end{document}